\documentclass[11pt, a4paper]{JHEP3}
\usepackage{amsmath, amssymb}
\usepackage{graphicx}
\def\be{\begin{eqnarray}}
 \def\ee{\end{eqnarray}}
 \def\0{\nonumber}
\def\cos{{\rm cos}}
\def\sin{{\rm sin}}

\def\be{\begin{eqnarray}}
 \def\ee{\end{eqnarray}}
 \def\0{\nonumber}
\def\cos{{\rm cos}}
\def\sin{{\rm sin}}

\def\sin{{\rm sin}}

\title{Marginal Deformations as Lower Dimensional D--brane Solutions in Open String Field theory}
\author{ Bum-Hoon Lee\\
Center for Quantum SpaceTime (CQUeST) and Department of Physics,\\
Sogang University, Shinsu-dong 1, Mapo-gu, Seoul, Korea\\
E-mail:  \email{bhl@sogang.ac.kr}}

\author{ Chanyong Park\\
Center for Quantum SpaceTime (CQUEeST), Sogang University\\
Shinsu-dong 1, Mapo-gu, Seoul, Korea\\
E-mail:  \email{cyong21@sogang.ac.kr}}

\author{D.D.Tolla\\
Center for Quantum SpaceTime (CQUeST), Sogang University\\
Shinsu-dong 1, Mapo-gu, Seoul, Korea\\
E-mail:  \email{tolla@sogang.ac.kr}}
 \abstract{By direct calculation we showed that a finite analytic solution for
 marginal deformation of open string field theory, by a matter primary operator
 with singular OPE, can be obtained to all orders in the deformation parameter. In particular, we obtained solutions
 that describe lower dimensional D--branes and our results agree with the results obtained when
 the same problem is treated in the world-sheet conformal field theory language.}
\begin{document}
\maketitle
\section{Introduction}
One point of view of understanding D-branes is that they are
solutions of string field theory equation of motion. Different
solutions of string field theory equation of motion represent
different two dimensional conformal field theory (CFT)
backgrounds.  Inspired by the Schnabl's  analytic construction of
open string field theory (OSFT) equation of motion representing
the tachyon vacuum  \cite{Schnabl05} (see
\cite{Okawa1,Ellwood:2006ba,RZ06,ORZ,Fuchs2,Fuchs0,Fuchs1,Bonora:2007tm,Erler:2006hw,Erler:2006ww,Imbimbo:2006tz}
for more development on this), recently several more solutions
have been obtained
\cite{Schnabl:2007az,KORZ,Fuchs3,Okawa:2007ri,Okawa3,Fuchs:2007gw,Kiermaier:2007ki}
for both bosonic and supersymmetric OSFT. These new solutions
describe conformal field theories that are deformed by exactly
marginal operators.  Among them are the solutions representing the
CFT of lower-dimensional D-branes.

It is a well established fact
\cite{Sen:1999mh,Recknagel:1998ih,Callan:1994ub,Polchinski:1994my,Callan:1993mw,Kogetsu:2004te}
that the boundary conformal field theory (BCFT) describing a
$D{p}$--brane in bosonic string theory is identical to that of a
$D(p+1)$-- brane deformed by an exactly marginal boundary operator.
More precisely, one can deform the former BCFT into the later by
adding an exactly marginal boundary term \be \delta
S_{ws}={\tilde\lambda}\int dt\sqrt{2}\cos(X(t)) \ee to the
world-sheet action, where $X$ is the direction transverse to the
$Dp$--brane, t is a coordinate on the world-sheet boundary and
${\tilde\lambda}$ is a free parameter. At
(${\tilde\lambda}=\pm{1\over 2}$) with this we obtain a periodic
array of $Dp$--branes, with Dirichlet boundary condition on $X$,
placed at ($x=(2n+1)\pi$) if we choose the plus sign and at
($x=2n\pi$) if we choose the minus sign.

An alternative description of marginal deformations in the
framework of string field theory was considered in
\cite{Sen:1990hh,Sen:2000hx,Takahashi:2002ez,Kluson:2002av,Kluson:2003xu,Sen:2004cq}.
It was shown that,  switching on a boundary marginal deformation
operator give rise to a string field theory configuration
corresponding to a new classical solution of the equation of
motion of OSFT formulated around the original undeformed BCFT. In
these investigations, mainly the level truncation method in the
Siegel gauge was used and switching on the marginal boundary
operator in world sheet was interpreted  as giving vacuum
expectation values (vev) to the fields associated to the tachyonic
and the massless open string modes. The rolling tachyon solution
by Sen \cite{Sen:2002nu} is the best example for such description
of marginal deformations in  string field theory framework, where
the vev was turned on only for the tachyonic mode.

A recent construction of analytic solutions for marginal
deformations in OSFT \cite{Schnabl:2007az,KORZ} used the recursive
technique developed in  \cite{Sen:2002nu} in a new gauge
introduced by Schnabl in \cite{Schnabl05} ($B\Psi=0$), where $B$
is the antighost zero mode in the conformal frame of the sliver.
The ansatz for the solutions were given by a series expansion in
some parameter $\lambda$ which to the first order can be
identified with the coupling constant ${\tilde\lambda}$ of the
exactly  marginal operator we mentioned above. One can then solve
the equation of motion at each order of $\lambda$. These
techniques were very effective to obtain solutions generated by a
marginal deformation operator $V(z)$ that has  a regular OPE with
itself. When the OPE is singular, divergences arises as the
separation between boundary insertions approaches zero and one
needs to add counter terms at each order of $\lambda$ to
regularize it. However, the form of the counter terms were
obtained only up to the third order by a clever guess and  their
forms for higher order terms are not known. One purpose of this
paper is to study the origin of the divergences in the case of
marginal deformations with singular OPE and to develop a method to
determine the counter terms necessary to cancel the divergences at
any order. In an earlier work \cite{KORZ}it was mensioned as an open propblem that some of the counter terms violate the gauge condition even though the solutions were constracted to respect the gauge. In this paper we will demonstrate by explicite calculatios that in the case of singular OPE marginal deformation, unlike the regular ones, only a piece of the solution can respect the Schnabl gauge and it is not surprising to have counter terms out side the gauge.

The rest of the paper is organized as follows. In section 2 we
will consider solutions with both regular OPE and singular OPE. We
will show  that the main difference between these solution is the
fact that the first one can be expanded only in terms of states
with positive $L$ eigenvalue, where $L$ is the Virasoro operator
$L_0$ in the conformal frame of the sliver, while the second one
contains the eigenvalues $0$ and $-1$ as well. In the same section
we will show that the divergences in the case of singular OPE
arise from inverting the $L$ operator on zero eigenvalue states
and using the Schwinger representation of ($L^{-1})$ on negative
eigenvalue states. Knowing the origin of the divergences we could
easily determine the form of the counter terms to add at each
level to regularize the solution. In section 3 we will use the
procedure we developed in section 2 to write solution representing
array of D24--branes, which are obtained when an exactly marginal
boundary deformation is turned on along the 25--th direction. In
section 4 we will discuss our results.

\section {The action of B/L and the OPE of V}
The linearized string field theory equation of motion ($Q_B\Psi=0$) is satisfied
by the state $\Psi^{(1)}=cV(0)|0\rangle$ corresponding to the operator $cV(0)$,
for any dimension one matter primary operator $V$. An ansatz for new class of solutions
for the non-linear equation of motion ($Q_B\Psi+\Psi\ast\Psi=0$) were made as an expansion in some parameter $\lambda$.
\be
\Psi_{\lambda}=\sum_{n=1}^{\infty}\lambda^{n}\Psi^{(n)},
\ee
with $\Psi^{(n)}$ satisfying
\be
Q_B\Psi^{(n)}=\Phi^{(n)},~~~~~~n>1 \label{eom}
\ee
where $\Phi^{(n)}$ is BRST exact and is given by
\be
\Phi^{(n)}=-\sum_{k=1}^{n-1}\Psi^{(n-k)}\ast\Psi^{(k)}\label{phin}
\ee

 If $\Psi$ is in Schnabl gauge ( $B\Psi_\lambda=0$) and there is no overlap between $\Phi^{(n)}$ and the kernel of $L$ the solution can be written as
\be \Psi^{(n)}={B\over L}\Phi^{(n)}\label{psin} \ee Further more
if $\Phi^{(n)}$ does not contain states with negative $L$
eigenvalues we can write \be
\Psi^{(n)}=\int_{0}^{\infty}dT~Be^{-TL}\Phi^{(n)}\label{sol1} \ee
In this section we will show that such a solution is allowed only
when $V(z)$ has a regular OPE with itself while in the case of
singular OPE, only part of the solution can be written as in
(\ref{sol1}). Here we notice that if not for  the action of
$L^{-1}$, in \ref{psin} the operators are inserted at finite
distances from each other along the real axis of the conformal
frame of the sliver and every thing  is regular, even if the
matter primary operator has a singular OPE with itself. However,
the action of $L^{-1}$ deletes a strip of certain width and make
the operators to collide. Therefore, the origin of any singularity
is the action of $L^{-1}$ on states of zero
 $L$ eigenvalues or its Schwinger representation on states of negative $L$
eigenvalues. We will see this in detail below.

 Lets begin with the regular OPE case where
\be
\lim_{z_1\to z_2}V(z_1)V(z_2)={\rm regular}
\ee
Using this we can easily verify that the commutation relation for the modes of V is
\be
[V_m,V_n]=\oint{dz_2\over 2\pi i}{\rm Res}_{z_1\to z_2}z_1^m z_2^n V(z_1)V(z_2)=0,~~~~~~\forall~ m,n
\ee
It is also true that for $m\ge 0$, $V_m|0\rangle=0$, as the conformal dimension of $V$ is one. We start our computation with the lowest level of  (\ref{phin}).
\be
\Phi^{(2)}=-\Psi^{(1)}\ast\Psi^{(1)}=-cV(0)|0\rangle\ast cV(0)|0\rangle
\ee
 In the conformal frame of the sliver
\be \Phi^{(2)}=-{\tilde c}{\tilde V}(0)|0\rangle\ast {\tilde
c}{\tilde V}(0)|0\rangle.\label{phi2} \ee Note that as $cV$ is a
primary operator of conformal dimension zero so that there is no
associated conformal factor infront. This star product can be
easily be carried out as it is the simplest case of the star
product of wedge states with insertions \be
U_r^\dagger U_r{\tilde\phi_1}(z_1)|0\rangle\ast U_s^\dagger U_s{\tilde\phi_2}(z_2)|0\rangle=U_{r+s-1}^\dagger U_{r+s-1}{\tilde\phi_1}(z_1+{s-1\over 2}){\tilde\phi_2}(z_2-{r-1\over 2})|0\rangle\0\\
\label{star1}
\ee
which we can write, after obvious shift of coordinate (${\tilde z}_i\to{\tilde z}_i+{r-1\over 2}$), as
\be
U_r^\dagger U_r{\tilde\phi_1}(z_1)|0\rangle\ast U_s^\dagger U_s{\tilde\phi_2}(z_2)|0\rangle=U_{r+s-1}^\dagger U_{r+s-1}{\tilde\phi_1}(z_1+{r+s-2\over 2}){\tilde\phi_2}(z_2)|0\rangle\0\\
\label{starprod} \ee where $U_r^\dagger U_r=e^{-{1\over
2}(r-2)L^+}$ with $L^+=L+L^\dagger$. If we have more than two
states to star multiply we use (\ref{star1}) associatively and do
the appropriate shift of coordinate at the end, as the shift we
have just made is not associative. In our simple case, which is
($r=s=2)$ we find
\be \Phi^{(2)}=-U_3^\dagger U_3{\tilde c}{\tilde
V}\left(1\right) {\tilde c}{\tilde V}\left(0\right)|0\rangle. \ee
Expanding both ${\tilde c}(z)$ and ${\tilde V}(z)$ in their
oscillator modes we can write \be \Phi^{(2)}=-e^{-{1\over
2}L^+}\sum_{l}\sum_{m}{\tilde c}_l{\tilde c}_{1}{\tilde V}_m
{\tilde V}_{-1} |0\rangle.
\ee
As  all commutations and
anticommutations of the oscillator modes appearing in this
expression are trivial the range of the indices will be
\be
\Phi^{(2)}=-\sum_{r=0}^{\infty}{1\over (-2)^r
r!}(L^+)^r\sum_{l=-\infty}^{1}\sum_{m=-\infty}^{-1}{\tilde
c}_l{\tilde c}_{1}{\tilde V}_m {\tilde V}_{-1}
|0\rangle.\label{phi2r}
\ee
Here we notice that each term of this
multiple sum is an eigenstate of $L$ with eigenvalue
($l_0=r-(l+m)\ge 1$) for ($r,l,m$) in the these ranges. Therefore,
we conclude that if V has a regular OPE with itself, there is no
overlap between the kernel of $L$ and $\Phi^{(2)}$  does not contain any term with negative $L$ eigenvalue.  For higher order $\Phi^{(n)}$ we will have similar
expression with more ${\tilde V}_m, {\tilde c}_l$ and
$B^+=B+B^\dagger$ insertions. With $l,m$ still in the range given
above and noting that $B^+$ raise the $L$ eigenvalue by one we see
that higher order $\Phi^{(n)}$ also does not contain negative or
zero $L$ eigenvalues.   Therefore, it is safe to invert $L$ or use
the Schwinger representation of $L^{-1}$ on $\Phi^{(n)}$ for
$\forall n>1$ when V has regular OPE with itself.

Next lets consider the case where $V$ has singular OPE, in particular
\be
V(z_1)V(z_2)={1\over (z_1-z_2)^2}+{\rm regular}.\label{OPE}
\ee
The commutation relation and the action on the vacuum of the oscillator modes will be
\be
 [V_m,V_n] =m\delta_{m,-n},~~~~~~~~~~ V_l|0\rangle=0,~~\forall~l\ge 0.
\ee Therefore, unlike the case in equation (\ref{phi2r}) we can
not drop all the positive modes of V and hence  $\Phi^{(2)}$ is
written as \be \Phi^{(2)}=-\sum_{r=0}^{\infty}{1\over (-2)^r
r!}(L^+)^r\sum_{l=-\infty}^{1}\sum_{m=-\infty}^{1}{\tilde
c}_l{\tilde c}_1{\tilde V}_m {\tilde V}_{-1}
|0\rangle\label{phi2ir} \ee or \be
\Phi^{(2)}&=&-\sum_{r=0}^{\infty}{1\over (-2)^r r!}(L^+)^r\sum_{l=-\infty}^{1}\sum_{m=-\infty}^{-1}{\tilde c}_l{\tilde c}_1{\tilde V}_m {\tilde V}_{-1} |0\rangle\0\\
&-&\sum_{r=0}^{\infty}{1\over (-2)^r
r!}(L^+)^r\sum_{l=-\infty}^{1}{\tilde c}_l{\tilde
c}_1|0\rangle.\label{phi2ir2} \ee The first line is exactly what
we have in the case of regular OPE and hence there is no ($l_0\le
0$) state in the first line. The $L$ eigenvalue of each term in
the second line is ($l_0=r-(l+1)\ge -1$). Therefore, in this case
there is an overlap between the kernel of $L$ and $\Phi^{(2)}$ and
it contains negative $L$ eigenvalue terms as well. The only
choices which give ($l_0=0$) are \be (r=0,l=-1),~~~~ (r=1,l=0) \ee
and the only one which gives ($l_0=-1$) is \be (r=0,l=0) \ee The
$(r=0,l=-1)$ case is ruled out by twist symmetry \cite{Schnabl05}, therefore,
$\Phi^{(2)}$ can be written as \be
\Phi^{(2)}&=&-\sum_{r=0}^{\infty}{1\over (-2)^r r!}(L^+)^r\sum_{l=-\infty}^{1}\sum_{m=-\infty}^{-1}{\tilde c}_l{\tilde c}_1{\tilde V}_m {\tilde V}_{-1} |0\rangle\0\\
&-&\sum_{r'}{1\over (-2)^{r'} r'!}(L^+)^{r'}\sum_{l'}{\tilde c}_{l'}{\tilde c}_{1}|0\rangle\0\\
&+&\left(-{\tilde c}_0{\tilde c}_1|0\rangle+{1\over 2}L^+{\tilde c}_0{\tilde c}_1|0\rangle\right)
\ee
where the primed indices are the corresponding unprimed indices without the cases which give ($l_0=0$) or ($l_0=-1$). The last line is BRST exact so that we can write
\be
\Phi^{(2)}&=&-\sum_{r=0}^{\infty}{1\over (-2)^r r!}(L^+)^r\sum_{l=-\infty}^{1}\sum_{m=-\infty}^{-1}{\tilde c}_l{\tilde c}_1{\tilde V}_m {\tilde V}_{-1} |0\rangle\0\\
&-&\sum_{r'}{1\over (-2)^{r'} r'!}(L^+)^{r'}\sum_{l'}{\tilde c}_{l'}{\tilde c}_{1}|0\rangle\0\\
&+& Q_B\left({\tilde c}_1|0\rangle-{1\over 2}L^+{\tilde c}_1|0\rangle\right)\0\\
&=& Q_B\left({\tilde c}_1|0\rangle-{1\over 2}L^+{\tilde
c}_1|0\rangle\right)+\Phi^{(2)}_{>} \ee where $\Phi^{(2)}_{>}$
contains only $l_0>0$ states. Up to some $Q_B$ closed term
$\Psi^{(2)}$ is given by \be \Psi^{(2)}={\tilde
c}_1|0\rangle-{1\over 2}L^+{\tilde c}_1|0\rangle+\Psi^{(2)}_{>}
\ee where $\Psi^{(2)}_{>}$ satisfies
$Q_B\Psi^{(2)}_{>}=\Phi^{(2)}_{>}$. Assuming $\Psi^{(2)}_{>}$ is
in the Schnabl gauge we can write \be
\Psi^{(2)}&=&{\tilde c}_1|0\rangle-{1\over 2}L^+{\tilde c}_1|0\rangle+\int_{0}^{\infty}dT~Be^{-TL}\Phi^{(2)}_{>}\0\\
&=&{\tilde c}_1|0\rangle-{1\over 2}L^+{\tilde c}_1|0\rangle+\lim_{\Lambda\to\infty}\int_{0}^{\Lambda}dT~Be^{-TL}[\Phi^{(2)}+\left({\tilde c}_0{\tilde c}_1|0\rangle-{1\over 2}L^+{\tilde c}_0{\tilde c}_1|0\rangle\right)]\0
\ee
Replacing $\int_{0}^{\Lambda}dT~Be^{-TL}$ by ${B\over L}-e^{-\Lambda L}{B\over L}$ on the terms with $l_0=0$ and $l_0=-1$ we obtain
\be
\Psi^{(2)}=\lim_{\Lambda\to\infty}\left(\int_{0}^{\Lambda}dT~Be^{-TL}\Phi^{(2)}+e^\Lambda{\tilde c}_1|0\rangle-{1\over 2}\Lambda\left[L^+{\tilde c}_1|0\rangle+B^+{\tilde c}_0{\tilde c}_1|0\rangle\right]-{1\over 2}L^+{\tilde c}_1|0\rangle\right)\0\\
=\lim_{\Lambda\to\infty}\left(-{1\over 2}(\Lambda+1)L^+{\tilde c}_1|0\rangle-{1\over 2}\Lambda B^+{\tilde c}_0{\tilde c}_1|0\rangle+e^\Lambda{\tilde c}_1|0\rangle-\int_{e^{-\Lambda}}^{1}dt~~\Psi^{(1)}\ast U_t^\dagger U_t|0\rangle\ast B_L^+\Psi^{(1)}
\right)\0\\
\label{psi2} \ee This solution is obtained by inverting $L$ only
on the positive eigenvalue terms of $\Phi^{(2)}$ so that it is
regular. We also notice that one can not fit the entire
$\Psi^{(2)}$ into the Schnabl gauge, only the portion which is
related to the positive  eigenvalue terms of $\Phi^{(2)}$ can
satisfy the gauge condition. The fact that this solution is
regular will be apparent when we will use it to calculate the
tachyon profile of the array of D24--brane solutions in the next
section.

Now we use the identity (see \cite{Schnabl05})
\be
\phi_1\ast B_L^+\phi_2=(-1)^{\phi_1}B_L^+(\phi_1\ast\phi_2)-(-1)^{\phi_1}(B_1\phi_1)\ast\phi_2 \label{idt}
\ee
and the fact that $B_1U_t^\dagger U_t|0\rangle=0$ and write $\Psi^{(2)}$ as
\be
\Psi^{(2)}&=&\lim_{\Lambda\to\infty}\left(\int_{e^{-\Lambda}}^{1}dt~~\{ B_L^+[\Psi^{(1)}\ast U_t^\dagger U_t|0\rangle\ast \Psi^{(1)}]-[B_1\Psi^{(1)}]\ast U_t^\dagger U_t|0\rangle\ast \Psi^{(1)}\}\right.\0\\
&+&\left.{\over} e^\Lambda{\tilde c}_1|0\rangle-{1\over 2}(\Lambda+1)L^+{\tilde c}_1|0\rangle-{1\over 2}\Lambda B^+{\tilde c}_0{\tilde c}_1|0\rangle\right).
\ee
This last form is convenient to calculate $\Phi^{(3)}$ which is given by
\be
\Phi^{(3)}=-\Psi^{(1)}\ast\Psi^{(2)}-\Psi^{(2)}\ast\Psi^{(1)}.
\ee
With the help of the identity (\ref{idt}) again, we obtain
\be
\Phi^{(3)}&=&\lim_{\Lambda\to\infty}\left(\int_{e^{-\Lambda}}^{1}dt~~ \{B_L^+[\Psi^{(1)}\ast\Psi^{(1)}\ast U_t^\dagger U_t|0\rangle\ast \Psi^{(1)}]-[B_1\Psi^{(1)}]\ast\Psi^{(1)}\ast U_t^\dagger U_t|0\rangle\ast \Psi^{(1)}\right.\0\\
&+&\left.\Psi^{(1)}\ast [B_1\Psi^{(1)}]\ast U_t^\dagger U_t|0\rangle\ast \Psi^{(1)}\}-e^\Lambda\Psi^{(1)}\ast{\tilde c}_1|0\rangle+{1\over 2}\Psi^{(1)}\ast L^+{\tilde c}_1|0\rangle\right.\0 \\
&-&\left.{1\over 2}\Lambda Q_B[\Psi^{(1)}\ast B^+{\tilde c}_1|0\rangle ]\right)\0\\
&-&\lim_{\Lambda\to\infty}\left(\int_{e^{-\Lambda}}^{1}dt~~ \{B_L^+[\Psi^{(1)}\ast U_t^\dagger U_t|0\rangle\ast \Psi^{(1)}\ast\Psi^{(1)}]-[B_1\Psi^{(1)}]\ast U_t^\dagger U_t|0\rangle\ast \Psi^{(1)}\ast\Psi^{(1)}\}\right.\0\\
&+&\left.{\over} e^\Lambda{\tilde c}_1|0\rangle \ast \Psi^{(1)}-{1\over 2}L^+{\tilde c}_1|0\rangle \ast \Psi^{(1)}-{1\over 2}\Lambda Q_B[B^+{\tilde c}_1|0\rangle\ast\Psi^{(1)}]\right).\label{Phi3}
\ee
Now we will need the following re-writings
\be
%\phi_1\ast B^+\phi_2&=&(-1)^{\phi_1}B^+(\phi_1\ast\phi_2)-(-1)^{\phi_1}(B_1\phi_1)\ast\phi_2 \0\\
%B^+\phi_1\ast \phi_2&=&B^+(\phi_1\ast\phi_2)+(-1)^{\phi_1}(\phi_1)\ast B_1\phi_2 \0\\
%L^+\phi_1\ast \phi_2&=&L^+(\phi_1\ast\phi_2)-\phi_1\ast K_1\phi_2 \0\\
%\phi_1\ast L^+\phi_2&=&L^+(\phi_1\ast\phi_2)+K_1\phi_1\ast \phi_2 \label{Lplus}
L^+\phi_1\ast \phi_2=-2{\partial\over\partial s} U_s^\dagger U_s\phi_1\ast \phi_2\left|_{s=2}\right.\0\\
\phi_1\ast L^+\phi_2=-2{\partial\over\partial s}\phi_1\ast  U_s^\dagger U_s\phi_2\left|_{s=2}\right..
\ee
Therefore,
\be
\Phi^{(3)}&=&\lim_{\Lambda\to\infty}\left(\int_{e^{-\Lambda}}^{1}dt~~ \{B_L^+[\Psi^{(1)}\ast\Psi^{(1)}\ast U_t^\dagger U_t|0\rangle\ast \Psi^{(1)}]-[B_1\Psi^{(1)}]\ast\Psi^{(1)}\ast U_t^\dagger U_t|0\rangle\ast \Psi^{(1)}\right.\0\\
&+&\left.\Psi^{(1)}\ast [B_1\Psi^{(1)}]\ast U_t^\dagger U_t|0\rangle\ast \Psi^{(1)}\}-{\over} e^\Lambda\Psi^{(1)}\ast{\tilde c}_1|0\rangle-{\partial\over\partial s}[\Psi^{(1)}\ast U_s^\dagger U_s{\tilde c}_1|0\rangle]\left|_{s=2}\right.\right.\0 \\
&-&\left.{1\over 2}\Lambda Q_B[ \Psi^{(1)}\ast B^+{\tilde c}_1|0\rangle] {\over}\right)\0\\
&-&\lim_{\Lambda\to\infty}\left(\int_{e^{-\Lambda}}^{1}dt~~ \{B_L^+[\Psi^{(1)}\ast U_t^\dagger U_t|0\rangle\ast \Psi^{(1)}\ast\Psi^{(1)}]-[B_1\Psi^{(1)}]\ast U_t^\dagger U_t|0\rangle\ast \Psi^{(1)}\ast\Psi^{(1)}\}\right.\0\\
&+&{\over} e^\Lambda{\tilde c}_1|0\rangle \ast \Psi^{(1)}+{\partial\over\partial s}[ U_s^\dagger U_s{\tilde c}_1|0\rangle \ast \Psi^{(1)}]\left|_{s=2}\right.- \left.{1\over 2}\Lambda Q_B[B^+{\tilde c}_1|0\rangle\ast\Psi^{(1)}]{\over}\right)\label{Phi31}
\ee

Since $B_1\Psi^{(1)}=V(0)|0\rangle$  we can use the standard
formula for star product of wedge states with insertions to
perform the star product. As usual, our aim is to single out the
terms with negative or zero eigenvalues of $L$ so that we can use
the Schiwnger representation (\ref{sol1}) of $Q_B$ on the
remaining terms of $\Phi^{(3)}$ to obtain  $\Psi^{(3)}$. It can be
easily verified that  the $Q_B$ exact terms in $\Phi^{(3)}$ do not
contain $l_0\le 0$ term, therefore, we will leave these terms as
they are.

\be
\Phi^{(3)}&=&\lim_{\Lambda\to\infty}\left[\int_{e^{-\Lambda}}^{1}dt~~ \left\{B_L^+U_{t+3}^\dagger U_{t+3}{\tilde c}{\tilde V}\left({t+1}\right){\tilde c}{\tilde V}\left({t}\right){\tilde c}{\tilde V}\left(0\right)|0\rangle\right.\right.\0\\
&-&U_{t+3}^\dagger U_{t+3}{\tilde V}\left({t+1}\right){\tilde c}{\tilde V}\left({t}\right){\tilde c}{\tilde V}\left(0\right)|0\rangle\0\\
&+&\left.U_{t+3}^\dagger U_{t+3}{\tilde c}{\tilde V}\left({t+1}\right){\tilde V}\left({t}\right){\tilde c}{\tilde V}\left(0\right)|0\rangle\right\}\0\\
&-&e^{\Lambda}U_{3}^\dagger U_{3}{\tilde c}{\tilde V}\left({1}\right){\tilde c}\left({0}\right)|0\rangle+{1\over 2}U_{3}^\dagger U_{3}L^+{\tilde c}{\tilde V}\left({1}\right){\tilde c}\left({0}\right)|0\rangle\0\\
&-&\left.{1\over 2}U_{3}^\dagger U_{3}\partial({\tilde c}{\tilde V})\left({1}\right){\tilde c}\left({0}\right)|0\rangle-{1\over 2}\Lambda Q_B[ \Psi^{(1)}\ast B^+{\tilde c}_1|0\rangle]\right]\0\\
&-&\lim_{\Lambda\to\infty}\left[\int_{e^{-\Lambda}}^{1}dt~~ \left\{B_L^+U_{t+3}^\dagger U_{t+3}{\tilde c}{\tilde V}\left({t+1}\right){\tilde c}{\tilde V}\left({1}\right){\tilde c}{\tilde V}\left({0}\right)|0\rangle\right.\right.\0\\
&-&\left.U_{t+3}^\dagger U_{t+3}{\tilde V}\left({t+1}\right){\tilde c}{\tilde V}\left({1}\right){\tilde c}{\tilde V}\left({0}\right)|0\rangle\right\}\0\\
&+&e^{\Lambda}U_{3}^\dagger U_{3}{\tilde c}\left({1}\right){\tilde c}{\tilde V}\left({0}\right)|0\rangle-{1\over 2}U_{3}^\dagger U_{3}L^+{\tilde c}\left({1}\right){\tilde c}{\tilde V}\left({0}\right)|0\rangle\0\\
&+&\left.{1\over 2}U_{3}^\dagger U_{3}\partial{\tilde c}\left({1}\right){\tilde c}{\tilde V}\left({0}\right)|0\rangle-{1\over 2}\Lambda Q_B[B^+{\tilde c}_1|0\rangle\ast\Psi^{(1)}]\right]
\ee
As we did for $\Phi^{(2)}$, after expanding both ${\tilde c}$ and ${\tilde V}$ in modes and also expanding $U_s^\dagger U_s$ in power of $L^+$ we see that each term in the multiple summation is an eigenstate of $L$. We would like to focus on the terms which contain $l_0\le 0$ and which are not $Q_B$ exact. Here we see that the $e^\Lambda$, the  $(\partial c)V$ and $\partial c$ terms contain such states. It is also easy to see that some contribution comes  from the lines $2,3$ and $7$. Using the commutation relation for the $V$ modes we can separate these terms from the others so that
\be
\Phi^{(3)}&=&\lim_{\Lambda\to\infty}\left(\left[1-2e^\Lambda+\int_{e^{-\Lambda}}^{1}dt~~f(t)\right]{\tilde c}_0{\tilde c}_1{\tilde V}_{-1}|0\rangle+\Phi^{(3)}_{>}(\rm{non-exact})\right.\0\\
&-& \left.{1\over 2}\Lambda Q_B[\Psi^{(1)}\ast B^+{\tilde
c}_1|0\rangle]-B^+{\tilde
c}_1|0\rangle\ast\Psi^{(1)}]\right)\label{phi32} \ee where
$\Phi^{(3)}_{>}(\rm{non-exact})$ contains only terms with $l_0>0$
and are not $Q_B$ exact and \be f(t)=2+{2\over t^2}+{2\over
(1+t)^2}. \ee Here we notice that unlike the $\Psi^{(2)}$ case,
now we have $Q_B$ non--exact $l_0=0$ terms, therefore, we can not
tell apart every term with $l_0=0$ of $\Psi^{(3)}$. However, still
there is a piece of $(\ref{Phi3})$ which is $Q_B$ exact. It is
convenient to write $\Phi^{(3)}$ as \be
\Phi^{(3)}=-\lim_{\Lambda\to\infty}{1\over 2}\Lambda
Q_B\left(\Psi^{(1)}\ast B^+{\tilde c}_1|0\rangle-B^+{\tilde
c}_1|0\rangle\ast\Psi^{(1)}\right)+\Phi^{(3)}_{rest}.\label{conv}
\ee
 With this we can see that the most general $\Psi^{(3)}$, up to some $Q_B$ closed addition, is
\be
\Psi^{(3)}=-\lim_{\Lambda\to\infty}{1\over 2}\Lambda\left(\Psi^{(1)}\ast B^+{\tilde c}_1|0\rangle-B^+{\tilde c}_1|0\rangle\ast\Psi^{(1)}\right)+\Psi^{(3)}_{rest},
\ee

where $\Psi^{(3)}_{rest}$ is defined as
\be
Q_B\Psi^{(3)}_{rest}=\Phi^{(3)}_{rest}.
\ee
We assume that $\Psi^{(3)}_{rest}$ is in the Schnabl gauge, so we can formally put $\Psi^{(3)}$ as

\be
\Psi^{(3)}_{0}&=&-\lim_{\Lambda\to\infty}{1\over 2}\Lambda\left(\Psi^{(1)}\ast B^+{\tilde c}_1|0\rangle-B^+{\tilde c}_1|0\rangle\ast\Psi^{(1)}\right)\0\\
&&+\lim_{\Gamma\to\infty}\left(\int_{0}^{\Gamma}dT~Be^{-TL}\Phi^{(3)}_{rest}
\right) \ee This has $Q_B$ closed divergent term which arise from
some of the $l_0=0$ terms of $\Phi^{(3)}_{rest}$ and needs to be
regularized.  From (\ref{phi32}) it is not difficult to realize
that the regularized   $\Psi^{(3)}$ will be \be
\Psi^{(3)}_{reg}&=&-\lim_{\Lambda\to\infty}{1\over 2}\Lambda\left(\Psi^{(1)}\ast B^+{\tilde c}_1|0\rangle-B^+{\tilde c}_1|0\rangle\ast\Psi^{(1)}\right)\0\\
&+&\lim_{\Gamma\to\infty}\left(\int_{0}^{\Gamma}dT~Be^{-TL}\Phi^{(3)}_{rest}-\lim_{\Lambda\to\infty}\left[ -2e^\Lambda+\int_{e^{-\Lambda}}^{1}dt~~f(t)\right]\Gamma{\tilde c}_1{\tilde V}_{-1}|0\rangle\right)\0\\
\ee Note that the added counter terms are all $Q_B$ closed and are
also in the Schnabl gauge, therefore, they will not affect the
equation of motion as well as the gauge condition. From equations
(\ref{Phi3}) and (\ref{conv}) we can easily read
$\Phi^{(3)}_{rest}$  and we finally obtain

\be
\Psi^{(3)}_{reg}&=&-\lim_{\Lambda\to\infty}{1\over 2}\Lambda\left(\Psi^{(1)}\ast B^+{\tilde c}_1|0\rangle-B^+{\tilde c}_1|0\rangle\ast\Psi^{(1)}\right)\0\\
&+&\lim_{\Gamma\to\infty}\lim_{\Lambda\to\infty}\left\{-\left[-2e^\Lambda+\int_{e^{-\Lambda}}^{1}dt~~f(t)\right]\Gamma{\tilde c}_1{\tilde V}_{-1}|0\rangle\right.\0\\
&-&e^\Lambda\int_{e^{-\Gamma}}^{1}dt_2~~{1\over t_2}\left[\Psi^{(1)}\ast U_{t_2}^\dagger U_{t_2}|0\rangle\ast B_L^+{\tilde c}_1|0\rangle+{\tilde c}_1|0\rangle\ast U_{t_2}^\dagger U_{t_2}|0\rangle\ast B_L^+\Psi^{(1)}\right]\0\\
&+&{1\over 2}\int_{e^{-\Gamma}}^{1}dt_2\left({1\over t_2}\left[-\Psi^{(1)}\ast U_{t_2}^\dagger U_{t_2}|0\rangle\ast B^+{\tilde c}_1|0\rangle+B^+{\tilde c}_1|0\rangle\ast U_{t_2}^\dagger U_{t_2}|0\rangle\ast \Psi^{(1)}\right]\right.\0\\
&+&\left.\Psi^{(1)}\ast U_{t_2}^\dagger U_{t_2}|0\rangle\ast B_L^+L^+{\tilde c}_1|0\rangle+L^+{\tilde c}_1|0\rangle\ast U_{t_2}^\dagger U_{t_2}|0\rangle\ast B_L^+\Psi^{(1)}{\over}\right)\0\\
&+& \left.\int_{e^{-\Lambda}}^{1}dt_1\int_{e^{-\Gamma}}^{1}dt_2~t_2\left[\Psi^{(1)}\ast U_{t_2}^\dagger U_{t_2}|0\rangle\ast(-B_L^+)\Psi^{(1)}\ast U_{t_1t_2}^\dagger U_{t_1t_2}|0\rangle\ast(-B_L^+)\Psi^{(1)}\right]\right.\0\\
&+& \left.\int_{e^{-\Lambda}}^{1}dt_1\int_{e^{-\Gamma}}^{1}dt_2~t_2\left[\Psi^{(1)}\ast U_{t_1t_2}^\dagger U_{t_1t_2}|0\rangle\ast(-B_L^+)\Psi^{(1)}\ast U_{t_2}^\dagger U_{t_2}|0\rangle\ast(-B_L^+)\Psi^{(1)}\right]
\right\}\0\\
\label{Psi3} \ee In the third and the fourth lines it is clear
that the integrands are not well defined in the region $t_2\to 0$.
However, the singularities coming form this region are cancelled
partly by the corresponding counter terms in the second line, the
($-2e^\Lambda$) term and partly by similar divergences coming from
the last two lines. There are other divergences arising from the
last two line. These will be cancelled by the remaining part of
the counter term in the second line and there is no more
divergence related to $t_2\to  0$. Since the divergences related
to $t_1\to 0$ has already been regularized at level 2, this result
is perfectly regular. We will demonstrate this cancellation of
divergences in the next section using a particular example. We
would like to emphasize also,  like it is at level 2 here again
the entire solution can not be in the Schnabl gauge, only the part
which is obtained from the $Q_B$ non-exact piece of $\Phi^{(3)}$
is in the gauge.
\newpage
For level 4 calculation we would like to focus entirely on the terms which have zero or negative $L$ eigenvalues. Separating these terms from the rest we can write $\Phi^{(4)}$ as
\be
\Phi^{(4)}&=&-\left[\Psi^{(3)}\ast\Psi^{(1)}+\Psi^{(1)}\ast\Psi^{(3)}+\Psi^{(2)}\ast\Psi^{(2)}\right]_{>}\0\\
&+&\lim_{\Lambda\to\infty}\lim_{\Gamma\to\infty}\left\{\left[-\Lambda+\left(4e^{\Lambda}-2\int_{e^{-\Lambda}}^{1}dtf(t)\right)\Gamma
+e^{\Lambda}\int_{e^{-\Gamma}}^{1}dt_2\left(-{2\over t_2}-{2\over t_2(1+t_2)^2}\right)\right.\right.\0\\
&+&\int_{e^{-\Gamma}}^{1}dt_2\left(-{1\over t_2}-{1-t_2\over t_2(1+t_2)^2}\right)-\int_{e^{-\Gamma}}^{1}dt_2{t_2-3\over (1+t_2)^3}\0\\
&+&\int_{e^{-\Lambda}}^{1}dt_1\int_{e^{-\Gamma}}^{1}dt_2~t_2\left({2\over t_2^2(l_2+1)^2}+{2\over l_2^2(t_2+1)^2}+{2\over (l_2-t_2)^2}+t_2\to t_1t_2\right)\0\\
&-&\int_{e^{-\Lambda}}^{1}dt_1\int_{e^{-\Lambda}}^{1}dt_2\left({t_2\over (l_2'+1)^2}+{t_2\over (t_1+1)^2(t_2+1)^2}+{1\over t_1t_2^2}\right)\0\\
&+&\int_{e^{-\Lambda}}^{1}dt_1\int_{e^{-\Lambda}}^{1}dt_2\left({t_2+1\over (l_2'+1)^2}+{t_2+1\over t_1^2t_2^2}+{1\over (t_2+1)(t_1+1)^2}\right)\0\\
&-&e^{\Lambda}\left(-\int_{e^{-\Lambda}}^{1}dt_1{1\over t_1^2}+{e^{-\Lambda}\over 2}\int_{e^{-\Lambda}}^{1}dt_1{1\over t_1^2}-\int_{e^{-\Lambda}}^{1}dt_2{1\over t_2}+\int_{e^{-\Lambda}}^{1}dt_2{t_2+1\over t_2^2}-e^{\Lambda}-{(\Lambda+1)}\right)\0\\
&-&\left.{(\Lambda+1)\over 2}\int_{e^{-\Lambda}}^{1}dt_2{1\over t_2}-{\Lambda\over 2}\int_{e^{-\Lambda}}^{1}dt_2{1\over t_2^2}+{\Lambda(\Lambda+1)\over 4}\right]{\bf Q_BL^{+}{\tilde c}_1|0\rangle}\0\\
%%%%%%%%%%%%%%%%%%%%%%%%%%%%%%%%%%%%%%%%%%%%%%%%%%%%%%%%%%%%%%%
&+&\left[\Lambda+\left(-2e^{\Lambda}+\int_{e^{-\Lambda}}^{1}dtf(t)\right)\Gamma+e^{\Lambda}\int_{e^{-\Gamma}}^{1}dt_2\left({1+t_2\over t_2}+{1\over t_2(1+t_2)}\right)\right.\0\\
&+&\int_{e^{-\Gamma}}^{1}dt_2\left({1+t_2\over 2t_2}+{1-t_2\over 2t_2(1+t_2)}\right)+\int_{e^{-\Gamma}}^{1}dt_2\left( 1+{t_2-1\over 2(1+t_2)}\right)\0\\
&-&\int_{e^{-\Lambda}}^{1}dt_1\int_{e^{-\Gamma}}^{1}dt_2~t_2(l_2+1)\left({1\over t_2^2(l_2+1)^2}+{1\over l_2^2(t_2+1)^2}+{1\over (l_2-t_2)^2}+t_2\to t_1t_2\right)\0\\
&+&\int_{e^{-\Lambda}}^{1}dt_1\int_{e^{-\Lambda}}^{1}dt_2{l_2'+1\over 2}\left({t_2\over (l_2'+1)^2}+{t_2\over (t_1+1)^2(t_2+1)^2}+{1\over t_1t_2^2}\right)\0\\
&-&\int_{e^{-\Lambda}}^{1}dt_1\int_{e^{-\Lambda}}^{1}dt_2{l_2'+1\over 2}\left({t_2+1\over (l_2'+1)^2}+{t_2+1\over t_1^2t_2^2}+{1\over (t_2+1)(t_1+1)^2}\right)\0\\
&-&e^{\Lambda}\int_{e^{-\Lambda}}^{1}dt_1{t_1+1\over 2t_1^2}+{\Lambda+1\over 2}\int_{e^{-\Lambda}}^{1}dt_1{t_1+3\over 2t_1^2}-{\Lambda\over 2}\int_{e^{-\Lambda}}^{1}dt_1{t_1+1\over 2t_1^2}\0\\
&-&e^\Lambda\left(\int_{e^{-\Lambda}}^{1}dt_2\left({t_2+1\over 2t_2}-{(t_2+1)^2\over 2t_2^2}\right)+{e^{\Lambda}\over 2}+{3(\Lambda+1)\over 2}+e^{-\Lambda}{(\Lambda+1)\over 2}\int_{e^{-\Lambda}}^{1}dt_2{1\over t_2}\right)\0\\
&+&{3(\Lambda+1)\over 4}\int_{e^{-\Lambda}}^{1}dt_2{t_2+1\over t_2}-(\Lambda+1)^2+{\Lambda\over 2}\int_{e^{-\Lambda}}^{1}dt_2{t_2+1\over 2t_2^2}\0\\
&-&\left.\left.{3\Lambda(\Lambda+1)\over 8}\right]{\bf Q_BL^{+}{\tilde c}_1|0\rangle}\right\}
\ee
where $l_2=t_2+t_1t_2$ and $l_2'=t_1+t_2$. Since the terms in the square brackets are just numerical factors we may write $\Phi^{(4)}$ as
\be
\Phi^{(4)}=\Phi^{(4)}_>+A{\bf Q_B{\tilde c}_1|0\rangle}+B{\bf Q_BL^{+}{\tilde c}_1|0\rangle}.
\ee
We see that $\Phi^{(4)}$ has the same  form as  $\Phi^{(2)}$ such that the terms with zero or negative $L$ eigenvalues are $Q_B$ exact. Therefore, we follow the same procedure as in level two to solve for  $\Psi^{(4)}$. Up to some $Q_B$ closed term the anzats for $\Psi^{(4)}$ is
\be \Psi^{(4)}=A{\tilde
c}_1|0\rangle+BL^+{\tilde c}_1|0\rangle+\Psi^{(4)}_{>}
\ee where $\Psi^{(4)}_{>}$ satisfies
$Q_B\Psi^{(4)}_{>}=\Phi^{(4)}_{>}$. Assuming $\Psi^{(4)}_{>}$ is
in the Schnabl gauge we can write
\be
\Psi^{(4)}&=&A{\tilde c}_1|0\rangle+BL^+{\tilde c}_1|0\rangle+\int_{0}^{\infty}dT~Be^{-TL}\Phi^{(4)}_{>}
\ee
In obtaining this solution we applied $L^{-1}$ or its Schwinger form only on terms with positive definite $L$ eigenvalues. Therefore, at this level we didn't produce any new divergent term. Since we have already regularized all the lower level $\Psi^{(i)}$s'  it is clear that $\Psi^{(4)}$ is  regular. Like the lower levels we see that also at this level the solution contains a gauge condition violating term, which is a charcterstic of the solutions with singular OPE.

Now lets generalize our procedure to arbitrary level. By now it is clear that
divergences arise only when there are zero or negative $L$ eigenvalue terms in $\Phi^{(n)}$.
Noting that $\Phi^{(n)}$ has to be of ghost number two and has to be twist even, one can easily see that
the only terms which can appear in $\Phi^{(n)}$ and can have a negative or zero eigenvalue
are $\left({\tilde c}_0{\tilde c}_1|0\rangle,~~L^+{\tilde c}_0{\tilde c}_1|0\rangle,~~
{\tilde c}_0{\tilde c}_1{\tilde V}_{-1}|0\rangle\right)$, which are exactly what we have at levels two and three. In particular, if $n$ is even only the first two of these terms (which are $Q_B$ exact) appear. The reason is that  applying the commutation relation for each pair of $V$ kills all the $V$ operators and hence the term of the third kind can not appear.   In this case we follow the procedure of level two to solve for $\Psi^{(n)}$. For odd $n$, where we have odd number of $V$ operators and the pairing will leave one $V$, only the last term appears and it give rise to a new divergent term in  $\Psi^{(n)}$. However, this new divergent term is $Q_B$ closed as well as satisfy the Schnabl gauge so that we can subtruct it out to get a regular solution.  Therefore, our procedure can be used at any level.
\section{The tachyon profile}
In this section we consider the special case of a marginal
deformation corresponding to a periodic array of D24--branes. The
dimension one boundary matter primary operator $V(z)$ giving such
a solution is \be V(z)={1\over {\sqrt 2}}[V^+(z)+V^-(z)],~~~~~~
{\rm with}~~V^\pm=e^{\pm iX(z)} \ee where we choose $\alpha'=1$
and $X(z)=X^{25}(z)$. We can easily see that the OPE of $V$ with
itself is given by (\ref{OPE}) and hence it is an example of
singular OPE solutions we saw in section 2. Our aim in this
section is to calculate the $x$ dependence of the tachyon field
level by level and verify that the solution are regular and they
are indeed correspond to array of D24--branes. The calculation beyond the thrid level is too complicated so we restrict our treatement in this section to the first three levels. Actually, the over all shape of the tachyon profile does not change when we consider higher level contributions, what changes is the depth of its minima, to which we are not intending to associate any physical meaning for the reason we will give in the discussion section.

Since the result at level one is trivial we start with level two
calculations. At level two $x$ dependence of $\Psi^{(2)}$ must be
of the form \be
\Psi^{(2)}=\left(e^{2iX(0)}+e^{-2iX(0)}\right)\left[\beta_{2}^{2}c_1|0\rangle+...\right]+\left[\beta_{0}^{2}c_1|0\rangle+...\right].
\ee The dotes indicate higher level space-time fields and the
coefficients $\beta_n^{2}$ are given by \be
\beta_{n}^{2}=\langle\phi_{\pm n}, \Psi^{(2)}\rangle,~~~~~
\phi_{\pm n}=e^{\pm inX(0)}c\partial c(0)|0\rangle \ee where we
have ignored the irrelevant space time volume factor. By momentum
conservation $\beta_2^2$ gets a contribution only from the last
term of (\ref{psi2}) which is given by \be
\beta_{2}^{2}&=&{1\over 2}\left\langle\phi_{-2},\lim_{\Lambda\to\infty}\int_{e^{-\Lambda}}^{1}dt~~cV^+(0)|0\rangle\ast U_t^\dagger U_t|0\rangle\ast (-B_L^+)cV^{+}(0)|0\rangle\right\rangle\0\\
&=&{1\over
2}\left\langle\phi_{2},\lim_{\Lambda\to\infty}\int_{e^{-\Lambda}}^{1}dt~~cV^-(0)|0\rangle\ast
U_t^\dagger U_t|0\rangle\ast
(-B_L^+)cV^{-}(0)|0\rangle\right\rangle\0 \ee Each of the
$V^{\pm}$'s  gives the regular OPE solutions and the above result
have been calculated in \cite{KORZ}, and the answer is \be
\beta_2^2={1\over 2}(0.15206). \ee $\beta_0^2$ gets a contribution
from all the terms in (\ref{psi2}). Using the definitions \be
L^+=-2(K_1^L-K_1),~~~~~B^+=-2(B_1^L-B_1) \ee and noting that ($K_1
c_1|0\rangle+B_1c_0c_1|0\rangle=0$) we can rewrite $\Psi^{(2)}$ in
the following more convenient way: \be
\Psi^{(2)}=\lim_{\Lambda\to\infty}\left(\Lambda \psi_0'-{1\over 2}L^+{\tilde c}_1|0\rangle+ e^\Lambda{\tilde c}_1|0\rangle-\int_{e^{-\Lambda}}^{1}dt~~\Psi^{(1)}\ast U_t^\dagger U_t|0\rangle\ast B_L^+\Psi^{(1)}\right)\0\\
=\lim_{\Lambda\to\infty}\left(-\Lambda \psi_0'+e^\Lambda
\exp\left[-{e^{-\Lambda}L^+\over 2}\right]{\tilde
c}_1|0\rangle-\int_{e^{-\Lambda}}^{1}dt~~\Psi^{(1)}\ast
U_t^\dagger U_t|0\rangle\ast B_L^+\Psi^{(1)}\right)\0 \ee where
$\left(\psi_0'=K_1^Lc_1|0\rangle+B_1^Lc_0c_1|0\rangle\right)$ is
defined in \cite{Okawa1}.  With the help of the identity
($L^+=2L^+_L+K_1$), in the limit $\Lambda\to\infty$ it is not
difficult to show that \be \exp\left[-{e^{-\Lambda}L^+\over
2}\right]{\tilde c}_1|0\rangle=U_{e^{-\Lambda}+1}^\dagger
U_{e^{-\Lambda}+1}|0\rangle\ast\left({\tilde c}_1|0\rangle-{1\over
2}e^{-\Lambda}{\tilde c}_0|0\rangle\right) \ee so that \be
\Psi^{(2)}&=&\lim_{\Lambda\to\infty}\left[-\Lambda \psi_0'+e^\Lambda U_{e^{-\Lambda}+1}^\dagger U_{e^{-\Lambda}+1}|0\rangle\ast\left({\tilde c}_1|0\rangle -{1\over 2}e^{-\Lambda}{\tilde c}_0|0\rangle\right)\right.\0\\
&+&\left.\int_{e^{-\Lambda}}^{1}dt~~\Psi^{(1)}\ast U_t^\dagger U_t|0\rangle\ast(-B_L^+)\Psi^{(1)}\right].
\ee
Therefore,
\be
\beta_0^2&=&\lim_{\Lambda\to\infty}\left(-\Lambda\langle\phi_0, \psi_0'\rangle-e^\Lambda \left\langle f\circ\phi_0(0)\left({\tilde c}-{1\over 2}e^{-\Lambda}\partial{\tilde c}\right)(e^{-\Lambda}+1)\right\rangle_{e^{-\Lambda}+2}\right.\0\\
 &+&\left.\int_{e^{-\Lambda}}^{1}dt~~\left\langle f\circ\phi_0(0) {\tilde c}{\tilde V}^+(1){\cal B}{\tilde c}{\tilde V}^-(t+1)\right\rangle_{t+2}\right).\label{beta0}
\ee The subscripts indicate the width of the strip over which the
correlators are taken. Noting that $\phi_0=Q_Bc(0)|0\rangle$ we
have \be \langle\phi_0, \psi_0'\rangle=\langle c_{-1},
Q_B\psi_0'\rangle=0 \ee After a simple calculation the remaining
terms in the first line of (\ref{beta0}) gives \be e^\Lambda
\left\langle f\circ\phi_0(0)\left({\tilde c}-{1\over
2}e^{-\Lambda}\partial{\tilde
c}\right)(e^{-\Lambda}+1)\right\rangle_{e^{-\Lambda}+2}={2\over
\pi}(e^\Lambda+1)+{\cal O}(e^{-\Lambda}).\label{beta22} \ee Note
that when we do the star product of wedge states with insertions
(eq. \ref{starprod}) we insert the operator of the last state in
the star product, first on the strip obtained by gluing together
the strips of the individual state. This operator ordering is
opposite to the one we use when we calculate the correlator in
(\ref{beta0}) and as a result we got an extra minus sign. The
ghost part of the last line of (\ref{beta0}) have been calculated
in \cite{KORZ} and the matter part calculation is straight
forward. Finally, we  obtain \be
\int_{e^{-\Lambda}}^{1}dt~~\left\langle f\circ\phi_0(0) {\tilde c}{\tilde V}^+(1){\cal B}{\tilde c}{\tilde V}^-(t+1)\right\rangle_{t+2}&=&\int_{e^{-\Lambda}}^{1}dt~~{\pi\over t+2}\left[1-{2+t\over 2\pi}\sin\left({2\pi\over 2+t}\right)\right]\0\\
&\times&\sin^2\left({\pi\over 2+t}\right)\sin^{-2}\left({\pi t\over 2+t}\right)
\ee
Putting every thing together, and using Mathematica we obtain
\be
\beta_{0}^2=-{\sqrt{ 27}\over 4}=-1.29904
\ee
which is  regular as we anticipated.

To level 2 the tachyon profile is given by
\be
T(x)=-\cos(x)+(0.15206)\cos(2x)-1.29904
\ee
if we choose $\lambda=-1$ and
\be
T(x)=\cos(x)+(0.15206)\cos(2x)-1.29904
\ee
if we choose $ \lambda=+1$.\\
\begin{figure}[htbp]
    \hspace{-0.5cm}
\begin{center}
    \includegraphics[scale=1]{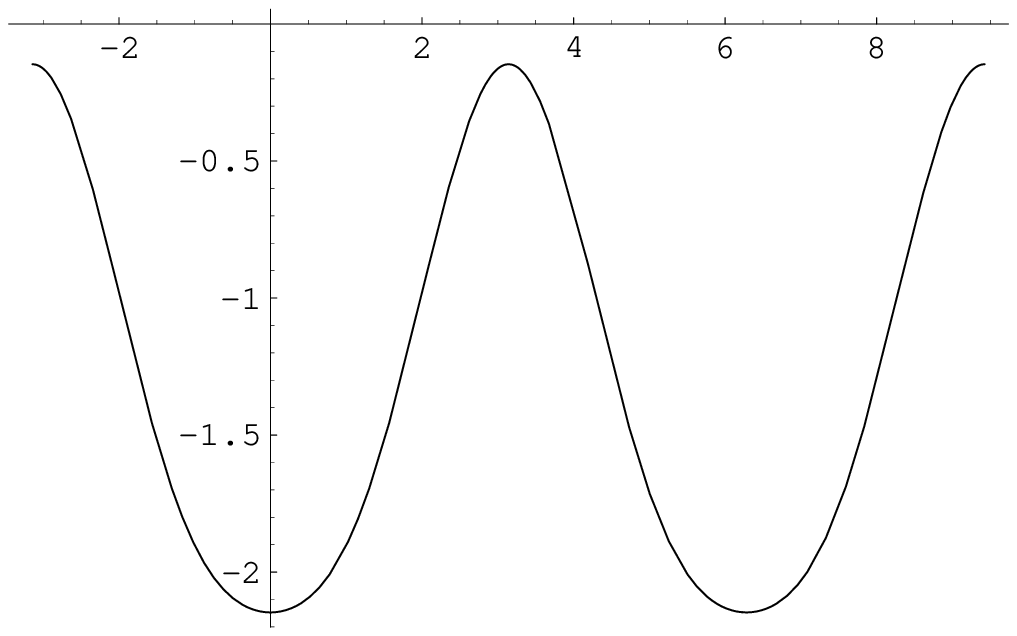}
    \end{center}
\caption{\emph{\small The level 2 approximation of the tachyon
profile for $\lambda=-1$}}
    \label{fig1}
\end{figure}
\begin{figure}[htbp]
    \hspace{-0.5cm}
\begin{center}
    \includegraphics[scale=1]{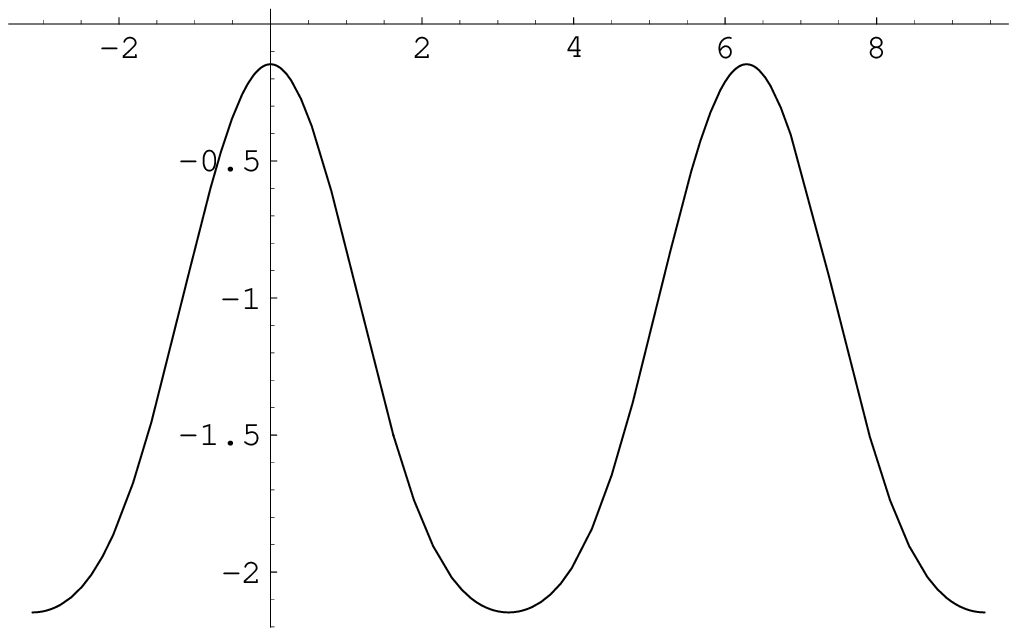}
    \end{center}
\caption{\emph{\small The level 2 approximation of the tachyon
profile for $\lambda=+1$ }}
    \label{fig2}
\end{figure}

Now lets proceed to level three calculations which should be of the form
\be
\Psi^{(3)}=\left(e^{3iX(0)}+e^{-3iX(0)}\right)\left[\beta_{3}^{3}c_1|0\rangle+...\right]+\left(e^{iX(0)}+e^{-iX(0)}\right)\left[\beta_{1}^{3}c_1|0\rangle+...\right]
\ee
with
\be
\beta_{n}^{3}=\langle\phi_{\pm n}, \Psi^{(3)}\rangle,~~~~~ \phi_{\pm n}=e^{\pm inX(0)}c\partial c(0)|0\rangle
\ee
By momentum conservation  only the last line of $(\ref{Psi3})$ matters in the calculation of $\beta_{3}^{3}$, which is given by
\be
\beta_{3}^{3}&=&{1\over \sqrt{8}}\left\langle\phi_{-3},\lim_{\Gamma\to\infty}\lim_{\Lambda\to\infty}\int_{e^{-\Lambda}}^{1}dt_1\int_{e^{-\Gamma}}^{1}dt_2~~cV^+(0)|0\rangle\ast U_{t_1}^\dagger U_{t_1}|0\rangle\right.\0\\
&&~~~~~~~~~~~~~~~~~\left.\ast (B_L^+)cV^{+}(0)|0\rangle\ast U_{t_2}^\dagger U_{t_2}|0\rangle\ast (B_L^+)cV^{+}(0)|0\rangle\right\rangle\0\\
&=&{1\over \sqrt{8}}\left\langle\phi_{3},\lim_{\Gamma\to\infty}\lim_{\Lambda\to\infty}\int_{e^{-\Lambda}}^{1}dt_1\int_{e^{-\Gamma}}^{1}dt_2~~cV^-(0)|0\rangle\ast U_{t_1}^\dagger U_{t_1}|0\rangle\right.\0\\
&&~~~~~~~~~~~~~\left.\ast (B_L^+)cV^{-}(0)|0\rangle\ast
U_{t_2}^\dagger U_{t_2}|0\rangle\ast
(B_L^+)cV^{-}(0)|0\rangle\right\rangle. \ee All the operator
involved have regular OPE and the result is that of \cite{KORZ}
again. \be \beta_{3}^{3}={1\over\sqrt{8}}(2.148\times10^{-3}) \ee
The calculation of $\beta_{1}^{3}$ is tedious but it is trivial,
it receives a contribution from all the terms in (\ref{Psi3}).
Next we list the contribution of each of them, where $l_i$ stands
for the contribution of the $i^{th}$ line in (\ref{Psi3}). \be
l_1=-\lim_{\Lambda\to\infty}{\Lambda\over\sqrt{2}}\left[{4\over
3}\left(1-{3\sqrt{3}\over 4\pi}\right)-1\right], \ee \be
l_2=\lim_{\Gamma\to\infty}\lim_{\Lambda\to\infty}\left[{\Gamma\over\sqrt{2}}\left(-2e^\Lambda+\int_{e^{-\Lambda}}^{1}dt~~f(t)\right)\right],
\ee \be
l_3=-\lim_{\Gamma\to\infty}\lim_{\Lambda\to\infty}{4e^\Lambda\over\sqrt{2}}\int_{e^{-\Gamma}}^{1}dt_2~~{1\over
t_2}\left\{{1\over t_2+2}\left[1-{2+t_2\over
2\pi}\sin\left({2\pi\over 2+t_2}\right)\right]\right\}, \ee \be
l_4=-\lim_{\Gamma\to\infty}{4\over\sqrt{2}}\int_{e^{-\Gamma}}^{1}dt_2~~{1\over
t_2}\left\{{1\over t_2+2}\left[1-{2+t_2\over
2\pi}\sin\left({2\pi\over 2+t_2}\right)\right]-{1\over 4}\right\},
\ee \be l_5={1\over\sqrt{8}}(0.734828) \ee \be
l_6&=&\lim_{\Gamma\to\infty}\lim_{\Lambda\to\infty}{2\pi^2\over\sqrt{8}}\int_{e^{-\Lambda}}^{1}dt_1\int_{e^{-\Gamma}}^{1}dt_2{t_2\over(2+l_2)^3}\left[1-{2+l_2\over 2\pi}\sin\left({2\pi\over 2+l_2}\right)\right] \0\\
&\times&\left[\sin^{-2}\left({\pi t_2\over 2+l_2}\right)\sin^{-2}\left({\pi (1+t_2)\over 2+l_2}\right)\sin^{-2}\left({2\pi\over 2+l_2}\right)\sin^{2}\left({\pi t_1t_2\over 2+l_2}\right)\sin^{2}\left({\pi\over 2+l_2}\right)\right.\0\\
&+&\sin^{-2}\left({\pi\over 2+l_2}\right)\sin^{-2}\left({\pi t_2\over 2+l_2}\right)\sin^{-2}\left({\pi t_1t_2\over 2+l_2}\right)\sin^{2}\left({\pi(1+t_2)\over 2+l_2}\right)\sin^{2}\left({2\pi\over 2+l_2}\right)\0\\
&+&\left.\sin^{-2}\left({\pi (1+t_2)\over 2+l_2}\right)\sin^{-2}\left({\pi t_1t_2\over 2+l_2}\right)\sin^{-2}\left({2\pi\over 2+l_2}\right)\sin^{2}\left({\pi t_2\over 2+l_2}\right)\sin^{2}\left({\pi\over 2+l_2}\right)\right]\0\\
\ee
\be
l_7&=&\lim_{\Gamma\to\infty}\lim_{\Lambda\to\infty}{2\pi^2\over\sqrt{8}}\int_{e^{-\Lambda}}^{1}dt_1\int_{e^{-\Gamma}}^{1}dt_2{t_2\over(2+l_2)^3}\left[1-{2+l_2\over 2\pi}\sin\left({2\pi\over 2+l_2}\right)\right] \0\\
&\times&\left[\sin^{-2}\left({\pi t_1t_2\over 2+l_2}\right)\sin^{-2}\left({\pi (1+t_1t_2)\over 2+l_2}\right)\sin^{-2}\left({2\pi\over 2+l_2}\right)\sin^{2}\left({\pi t_2\over 2+l_2}\right)\sin^{2}\left({\pi\over 2+l_2}\right)\right.\0\\
&+&\sin^{-2}\left({\pi\over 2+l_2}\right)\sin^{-2}\left({\pi t_1t_2\over 2+l_2}\right)\sin^{-2}\left({\pi t_2\over 2+l_2}\right)\sin^{2}\left({\pi(1+t_1t_2)\over 2+l_2}\right)\sin^{2}\left({2\pi\over 2+l_2}\right)\0\\
&+&\left.\sin^{-2}\left({\pi (1+t_1t_2)\over 2+l_2}\right)\sin^{-2}\left({\pi t_2\over 2+l_2}\right)\sin^{-2}\left({2\pi\over 2+l_2}\right)\sin^{2}\left({\pi t_1t_2\over 2+l_2}\right)\sin^{2}\left({\pi\over 2+l_2}\right)\right]\0\\
\ee
where $l_2=t_2+t_1t_2$. Note that we have made a change of sign on the first five terms for the same reason we gave after equation (\ref{beta22}). Each of these term can be evaluated numerically using mathematica and we finally obtain the following finite answer
\be
\beta_{1}^{3}=0.798956.
\ee
With this we can write the level three approximation of the tachyon profile as
\be
T(x)=-2.59791\cos(x)+(0.15206)\cos(2x)-1.29904-1.51887\times 10^{-3}\cos(3x)\0\\
\ee
for $\lambda=-1$ and
\be
T(x)=2.59791\cos(x)+(0.15206)\cos(2x)-1.29904+1.51887\times 10^{-3}\cos(3x)\0\\
\ee
for  $\lambda=+1$.\\
\begin{figure}[htbp]
    \hspace{-0.5cm}
\begin{center}
    \includegraphics[scale=1]{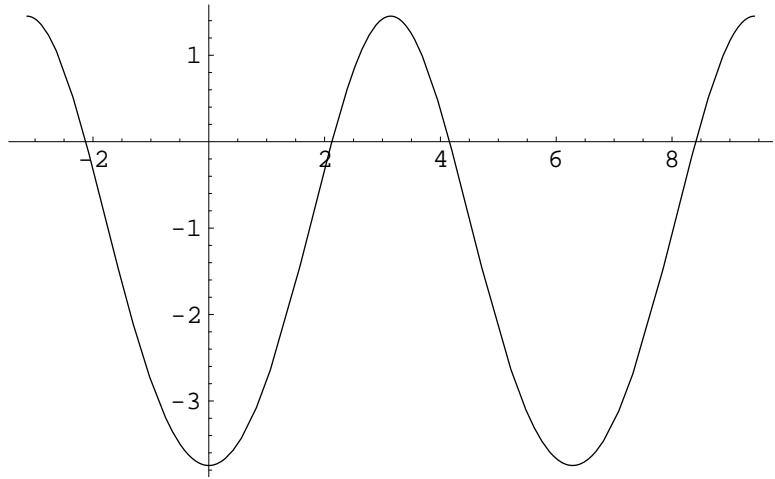}
    \end{center}
\caption{\emph{\small The level 3 approximation of the tachyon
profile for $\lambda=-1$ }}
    \label{fig3}
\end{figure}
\begin{figure}[htbp]
    \hspace{-0.5cm}
\begin{center}
    \includegraphics[scale=1]{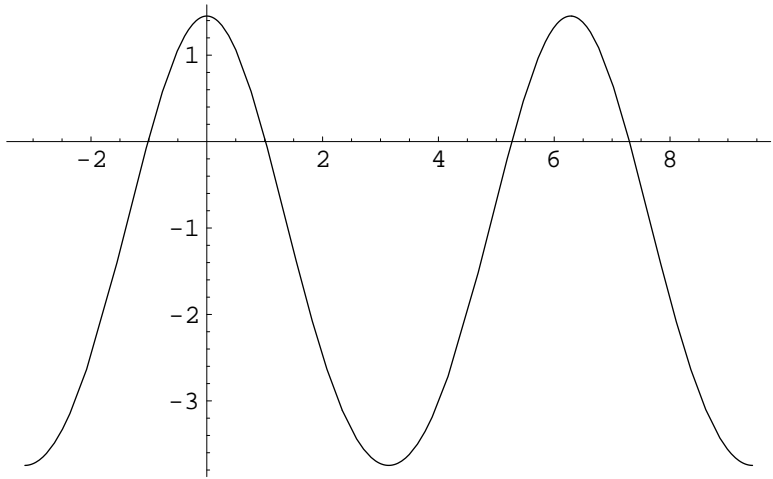}
    \end{center}
\caption{\emph{\small The level 3 approximation of the tachyon
profile for $\lambda=+1$ }}
    \label{fig4}
\end{figure}

At both level two an level three approximations, our result
confirms the results from conformal field theory description that
the $\cos(x)$ boundary deformation gives a solution representing a
periodic array of D--branes placed at odd integral multiple of
$\pi$ when the coupling ${\tilde\lambda}$ is positive and at even
integral multiple of  $\pi$ when the coupling ${\tilde\lambda}$ is
negative. In both cases the D-brane is situated at the minimum of
the interaction potential switched on along the boundary of the
world-sheet. To first order approximation, the  tachyon profile
and this interaction potential can be identified.  This means the first level approximation of the tachyon profile indicates the location of the D-branes. We just showed that including higher level contributions does not change the location of this minima and that means still with higher level contribution the tachyon profile minima is the location of the D-branes.
\section{Conclusion}
In this paper at the first place we could verify that an explicit
expansion of the $\Phi^{(n)}$  in terms of  definite $L$
eigenvalue states, contains zero and negative eigenvalues only
when the matter primary operator $V(z)$ has a singular OPE. This
fact helped us to identify the terms which give raise to
divergences in the case of singular OPE marginal deformations are
those with zero or negative eigenvalues. As these kind of terms
with the right ghost and twist number are very few, we conclude
that one can determine exactly the form of the counter terms which
have to be subtracted at any level of expansion in powers of
$\lambda$ to cancel the divergences associated with these terms.
We have also seen that unlike the regular OPE case, where
the entire solution satisfies the Schnabl gauge, only some piece
of the solution can satisfy the Schnabl gauge in the case of
singular OPE. We have shown this explicitely upto level 4 and it works the same  for levels higere than that as the the gauge violating terms of these levels are the same as those of the lowest levels.

In our computations we have considered only the case where the OPE
is given by $\ref{OPE}$. However, as what matters is the
commutation relation between the modes of $V(z)$, we believe that
the treatment in this paper can be generalized to any matter
primary operator with arbitrary singular OPE and hence different
commutation relation for the modes of $V(z)$.

In the second part of the paper we have considered the $\cos(x)$
marginal deformation, which from the world-sheet CFT point of
view, is known to represent a periodic array of D--branes located
at the minima of world-sheet potential. Using our results of the
first part we could calculate the tachyon profile up to level 3
and obtained a result which agrees with the world-sheet description.
Earlier, in the  string field theory framework, the tachyon profile
of a  lump solution have been obtained in \cite{Moeller:2000jy}
using the level truncation method, when the transverse direction
is compactified on a circle. Their result indicates that the lump
solution represents a single D--brane placed at $x=0$, which
coincides with our solution if we restrict our solution to one
period of the potential.

Lastly, we would like to comment on the depth of the minima of the
tachyon profile which seems to increase as we go to higher and
higher levels. As we mentioned before, to first order
approximation, the tachyon profile is related to the world-sheet
boundary interaction potential. This might lead one to the
conclusion that the depth of the minima of the tachyon profile is
related to the hight of the potential. However, this can not be
true since at each order approximation, our solutions are
determined up to a $Q_B$ closed additional terms, which if taken
into account will affects the depth of the minima of the tachyon
profile. Therefore, here we do not take the depth of the minima of
the tachyon profile seriously, all we need is its position which
is the position of the lower dimensional D--brane.

\acknowledgments

D.D.T. would like to thank Y. Okawa for his kind response to several
questions I have asked. C. Park would like to thank the Isaac Newton Institute for Mathematical Sciences, where I was visting while this work was in progress, for their hospitality. This work is supported by the Science Research Center Program of the Korean Science and Engineering Foundation through the Center for Quantum SpaceTime (CQUeST) of Sogang University with grant number R11-2005-021.

\end{document}